\documentstyle[prl,aps]{revtex}

\begin{document}

\draft
\twocolumn[
\widetext

\title{Hydrodynamic theory of an electron gas}
\author{I.~Tokatly\cite{MIET}, O.~Pankratov}
\address{Lehrstuhl f\"ur theoretische Festk\"orperphysik,\\
Universit\"at Erlangen-N\"urnberg, Staudtstr. 7/B2, 91054 Erlangen,
Germany}
\date{\today}
\maketitle

\widetext

\begin{abstract}
\leftskip 2cm 
\rightskip 2cm
The generalised hydrodynamic theory of an electron gas, which does not rely
on an assumption of a local equilibrium, is derived as the
long-wave limit of a kinetic equation. Apart from the common hydrodynamics
variables the theory includes the tensor fields of the higher moments of the
distribution function. In contrast to the Bloch hydrodynamics, the theory
leads to the correct plasmon 
dispersion and in the low frequency limit recovers the Navies-Stocks
hydrodynamics. The linear approximation to the generalised 
hydrodynamics is closely related to the theory of highly viscous fluids. 
\end{abstract}
 
 
]
\narrowtext
 
Hydrodynamic theory of an electron gas was heuristically introduced by Bloch
in 1933 \cite{Bloch} as an extension of Thomas-Fermi model. Only macroscopic
variables - electron density $n({\bf r},t)$, velocity ${\bf v}({\bf r},t)$,
pressure $P$, and electrostatic potential $\varphi ({\bf r},t),$ enter the
theory. The set of equations (continuity equation, Euler and Poisson
equations) becomes complete when the equation of state is added, and in the
original paper \cite{Bloch} Bloch identified $P$ with the kinetic pressure
of a degenerate Fermi gas. The Bloch's hydrodynamic theory (BHT) has been
applied to variety of kinetic problems 
\cite{Ritchie,Bennett,Fetter,Ying,Eguiluz,Barton,Halperin,Zaremba} with minor
improvements (inclusion of exchange,
correlation and quantum gradient corrections) \cite{Ying,Zaremba}.
 
From the microscopic point of view BHT cannot be a fully consistent theory
since it extends the collision-dominated hydrodynamics to the
electron gas where the collisionless (Vlasov) limit is most common. For
example, in the plasmon dispersion law 
$\omega ^{2}=\omega_{p}^{2}+v_{0}^{2}q^{2}$ BHT predicts for degenerate
electron gas 
$v_{0}^{2}=\frac{1}{3}v_{F}^{2}$ instead of a correct result 
$\frac{3}{5}v_{F}^{2}$ ($v_{F}$ is the Fermi velocity)  
\cite{Fetter,Eguiluz,Barton,Halperin,Lundqvist,Harris,Kleinman,Dobson,Jackson}
. At arbitrary degeneracy the
hydrodynamics gives $v_{0}^{2}=v_{s}^{2}$, where $v_{s}$ is 
velocity of sound, whereas in the kinetic theory $v_{0}^{2}$ equals to 
the mean square of the particle velocity $<v_{p}^{2}>$ \cite{Jackson}. It 
has been realized \cite{Lundqvist,Harris,Kleinman} that this 
discrepancy originates from the
assumption of a local equilibrium , which underlies the common 
hydrodynamic theory
\cite{Landau}. The assumption allows to reduce the kinetic equation for the
distribution function $f_{{\bf p}}({\bf r},t)$
\begin{equation}
\partial _{t}f_{{\bf p}}+\frac{{\bf p}}{m}\nabla f_{{\bf p}}+
e\nabla \varphi\frac{\partial f_{{\bf p}}}{\partial {\bf p}}=
I_{{\bf p}}\left[ f_{{\bf p}}\right]   
\label{1}
\end{equation}
to equations for macroscopic variables $n({\bf r},t)$,${\bf v}({\bf r},t)$
and, in general case, temperature $T({\bf r},t)$. The requirement of the
local equilibrium is fulfilled
if the characteristic time of the process $\tau \sim 1/\omega $ is much
longer than the inverse collision frequency $1/\nu $ and the typical length 
$L$ is greater than the mean free path $l\sim u/\nu $ ($u$ is the average
particle velocity)\cite{Klim} 
\begin{equation}
1/\tau \nu \sim \omega /\nu \ll 1,\qquad u/L\nu \ll 1.  \label{2}
\end{equation}
In a zero oder with respect to parameters (\ref{2}) Eq. (\ref{1})
reduces to $I_{{\bf p}}\left[ f_{{\bf p}}\right] $ $=0$ which means that 
$f_{{\bf p}}$ must have a locally equilibrium form. The stress tensor in the
co-moving (Lagrange) frame 
\begin{equation}
P_{ij}=\frac{1}{m}\sum_{{\bf p}}p_{i}p_{j}f_{{\bf p}+m{\bf v}},  \label{3}
\end{equation}
becomes diagonal $P_{ij}=P\delta _{ij}$ with $P$ being the local pressure.
As a result, the equations for the first three moments of the distribution
function i.e. density $n$, current ${\bf j}=n{\bf v}$ and stress tensor 
$P_{ij}$, form the closed set of hydrodynamics equations for an ideal liquid.
 
Due to the high frequency of the plasma waves at least the first of
inequalities (\ref{2}) is strongly violated and the tensor structure of 
$P_{ij}$ as well as that of higher moments becomes important. In a degenerate
case these tensors describe deformation of the Fermi sphere - the effect
completely ignored in the BHT. Within a linear
response theory \cite{Conti} this effect leads to the effective
shear modulus of a liquid. As we shall see below, this shear modulus is
directly related to the traceless part of $P_{ij}$. 
 
In this paper we present a generalised hydrodynamics which remains valid
under a strong violation of both conditions (\ref{2}). The theory
follows from the long-wave limit of the Boltzmann equation and requires
inclusion of the tensors of higher moments. We show that
this leads to correct plasmon dispersion.
 
The kinetic equation (\ref{1}) can be transformed in an infinite chain of
coupled equations for the moments of the distribution function\cite{Klim}.
For the first three moments: density $n$, current ${\bf j}=n{\bf v}$ and
stress tensor $P_{ij}$ we have 
\begin{eqnarray}
D_{t}n &+&n\nabla _{k}v_{k}=0  
\label{4} \\
mnD_{t}v_{i} &+&\nabla _{j}P_{ij}-en\nabla _{i}\varphi =0  
\label{5} \\
D_{t}P_{ij} &+&P_{ij}\nabla _{k}v_{k}+P_{ik}\nabla _{k}v_{j}+
P_{kj}\nabla_{k}v_{i}  
\label{6} \\
&+&\nabla _{k}Q_{ijk}=I_{ij},  \nonumber
\end{eqnarray}
where 
\begin{equation}
Q_{ijk}=\frac{1}{m^{2}}\sum_{{\bf p}}p_{i}p_{j}p_{k}f_{{\bf p}+m{\bf v}}
\label{7}
\end{equation}
is the third moment in the co-moving frame, 
$I_{ij}=\frac{1}{m}\sum_{{\bf p}}p_{i}p_{j}I_{{\bf p}}$ 
is the second moment of the collision integral and 
$D_{t}=\partial _{t}+{\bf v}\nabla$. In (\ref{4}-\ref{7}) the apparent
consequences of the conservation of number of particles, momentum and energy
in a collision process 
\begin{equation}
\sum_{{\bf p}}I_{{\bf p}}=0,\quad \sum_{{\bf p}}p_{i}I_{{\bf p}}=0,\quad
\sum_{{\bf p}}{\bf p}^{2}I_{{\bf p}}=0,  
\label{8}
\end{equation}
have been used. For a charged liquid one has to add the Poisson equation for
the scalar potential 
\begin{equation}
\nabla ^{2}\varphi =4\pi en-4\pi \rho _{ext},  
\label{9}
\end{equation}
where $\rho _{ext}$ is an external charge and $\varphi$ is an electrostatic
potential. The symmetric tensor $P_{ij}$ (\ref{3}) can be decomposed into 
the scalar and the traceless tensor part 
\begin{equation}
P_{ij}=P\delta _{ij}+\pi _{ij},\qquad Tr\pi _{ij}=0,  
\label{10}
\end{equation}
where $P=\frac{1}{3}TrP_{ij}$ is the pressure of an electron gas. 
 
Each equation in (\ref{4}-\ref{6}) is coupled to the next one through the
higher moment. If conditions (\ref{2}) are fulfilled, the higher moments 
are small and the infinite chain of equations can be decoupled 
(Chapman-Enskog or Grad methods, see \cite{Klim}). For example, as $\pi _{ij}$
and $Q_{ijk}$ vanish for locally equilibrium form of the distribution
function, they remain small under conditions (\ref{2}). The two 
tensors describe, respectively, viscosity and thermal conductivity in a
collision dominated liquid.
However, if conditions  (\ref{2}) are violated, $\pi _{ij}$ and $Q_{ijk}$  as
well as the higher-order moments are not small. Yet the infinite chain of
equations can still be decoupled if all physical quantities are slow varying
functions of ${\bf r}$. It is seen from Eq. (\ref{6}) that the third moment 
$Q_{ijk}$ enters only under a spatial derivative, hence it gives a correction
proportional to $1/L$. This remains true for all higher moments. 
Thus the truncation of the chain can be justified by
the smallness of the gradients of the moments instead of the smallness of
the moments itself. The dimensionless parameter of this expansion is 
\begin{equation}
\gamma \sim \frac{u}{Lmax\{\omega ,\nu \}}\ll 1.  
\label{11}
\end{equation}
It is clear that both inequalities (\ref{2}) can be violated while the
condition (\ref{11}) is fulfilled. 
 
To show the consistency of this decoupling procedure we consider 
equations for the moments
up to an arbitrary order in a transparent case of a one dimensional motion.
Generalisation for a three-dimensional case is straightforward, but
calculations are lengthy and will be published elsewhere.
 
Let the direction of the motion be along $x$-axis and hence 
$\nabla =\hat{{\bf x}}\partial _{x}$. Only diagonal $x$-components 
of all moments contribute
and the moment $M^{(k)}$ of the $k$-th order in the
laboratory frame is 
\begin{equation}
M^{(k)}=m^{1-k}\sum_{{\bf p}}p_{x}^{k}f_{{\bf p}}.\qquad k\ge 0.  
\label{12}
\end{equation}
To separate the macroscopic motion of liquid from the motion of particles we
introduce the relative moment $L^{(k)}$ i.e. the moment in the co-moving
frame 
\begin{equation}
L^{(k)}=m^{1-k}\sum_{{\bf p}}p_{x}^{k}f_{{\bf p}+m\hat{{\bf x}}v,}.
\label{13}
\end{equation}
which is related to $M^{(k)}$ as
\begin{equation}
M^{(k)}=\sum_{l=0}^{k}C_{k}^{l}v^{l}L^{(k-l)},  
\label{14}
\end{equation}
where $C_{k}^{l}=\frac{k!}{l!(k-l)!}$ are binomial coefficients. Apparently 
$M^{(0)}=mn$, $M^{(1)}=mnv$, $L^{(2)}=P_{xx}$,  $L^{(3)}=Q_{xxx}$ and 
$L^{(1)}\equiv 0$ by the definition.
 
For the $k$-th moment the Boltzmann equation gives
\begin{equation}
\partial _{t}M^{(k)}+\partial _{x}M^{(k+1)}-
k\frac{e}{m}M^{(k-1)}\partial_{x}\varphi =I^{(k)},  
\label{15}
\end{equation}
where $I^{(k)}$ is the $k$-th moment of the collision integral. According to
(\ref{8}), $I^{(0)}=I^{(1)}=0$. Substitution of (\ref{14}) in (\ref{15})
gives
\begin{eqnarray}
D_{t}L^{(0)} &+&L^{(0)}\partial _{x}v=0,  
\label{16} \\
L^{(0)}D_{t}v &+&\partial _{x}L^{(2)}-
\frac{e}{m}L^{(0)}\partial _{x}\varphi=0,  
\label{17} \\
D_{t}L^{(2)} &+&3L^{(2)}\partial _{x}v+\partial _{x}L^{(3)}=I^{(2)},
\label{18} \\
D_{t}L^{(k)} &+&(k+1)L^{(k)}\partial _{x}v+\partial _{x}L^{(k+1)}
\label{19} \\
&-&kL^{(k-1)}\frac{\partial _{x}L^{(2)}}{L^{(0)}}=I^{(k)},\quad k\ge 3.
\nonumber
\end{eqnarray}
Let us estimate the order of magnitude of the terms in Eq. (\ref{19}).
According to Eq.(\ref{16}), the spatial derivative $\partial _{x}v$ is of
the order of the inverse characteristic time $1/\tau \sim \omega $ hence the
first two terms in Eq. (\ref{19}) are proportional to $\omega $. The right
hand side of this equation is evidently proportional to the collision
frequency $\nu $. The last two terms in the left hand side of Eq. (\ref{19}) 
contain only spatial derivatives and are of the order of $u/L$. Thus
contribution of $L^{(k)}$ in the equation for $L^{(k-1)}$ contains additional
smallness $\gamma $ (\ref{11}), which means that the correction from 
$L^{(k)}$ to 
$L^{(0)}$ is of the order $\gamma ^{k}$. To obtain a theory valid up to 
$\gamma ^{k}$ one should keep $k+1$ equations for $L^{(l)}$ ($0\le l\le k$)
with the contribution of the $k+1$-th moment and the spatial derivative of
$L^{(2)}$ being omitted in the last equation. The resulting theory is a 
generalisation of hydrodynamics which is valid far from the equilibrium.
 
In a second order theory one should neglect $Q_{ijk}$ in Eq.(\ref{6}). Using
(\ref{10}) we obtain Eqs. (\ref{5}) and (\ref{6}) in terms of $P$ and $\pi
_{ij}$:
\begin{eqnarray}
&m&nD_{t}v_{i} +\nabla_{i}P+\nabla_{j}\pi_{ij}-en\nabla_{i}\varphi =0,
\label{20} \\
&D_{t}&P +\frac{5}{3}P\nabla_{k}v_{k}+\frac{2}{3}\pi_{ik}\nabla
_{k}v_{i}=0,  
\label{21} \\
&D_{t}&\pi_{ij} +\pi_{ij}\nabla_{k}v_{k}+\pi_{ik}\nabla_{k}v_{j}+\pi
_{kj}\nabla_{k}v_{i}-\frac{2}{3}\delta_{ij}\pi_{\mu k}\nabla
_{k}v_{\mu }  \nonumber \\
&+&P\left( \nabla_{i}v_{j}+\nabla_{j}v_{i}-\frac{2}{3}\delta_{ij}\nabla
_{k}v_{k}\right) =I_{ij},  
\label{22}
\end{eqnarray}
where Eq.(\ref{21}) is a trace of Eq.(\ref{6}) and Eq.(\ref{22}) is a
difference of (\ref{6}) and (\ref{21}). The continuity equation (\ref{4}),
equations (\ref{20}-\ref{22}) and Poisson equation (\ref{9}) are 
the generalised hydrodynamics equations which contain two scalars $n$ 
and $P$, one vector ${\bf v}$
and one traceless tensor $\pi _{ij}$. To make the system closed we also
need an expression for $I_{ij}$. However, in a charged liquid
contribution of $I_{ij}$ to
eigenmodes disappears since these modes (plasmons) lie in the
high frequency region $\omega /\nu \gg 1$  where the collisionless limit is
reached. In this limit the linearised equations are
\begin{eqnarray}
\partial _{t}\delta n &+&n_{0}\nabla_{k}v_{k}=0  
\label{23} \\
mn_{0}\partial_{t}v_{i} &+&\nabla_{i}\delta P+
\nabla_{j}\pi_{ij}-en_{0}\nabla_{i}\varphi =0  
\label{24} \\
\partial_{t}\delta P &+&\frac{5}{3}P_{0}\nabla_{k}v_{k}=0  
\label{25} \\
\partial_{t}\pi_{ij} &+&P_{0}\left( \nabla_{i}v_{j}+
\nabla_{j}v_{i}-\frac{2}{3}\delta_{ij}\nabla_{k}v_{k}\right) =0,  
\label{26}
\end{eqnarray}
where $\delta n$ and $\delta P$ are deviations from the equilibrium density 
$n_{0}$ and pressure $P_{0}$. To obtain the plasmon dispersion we 
consider a plane-wave solution $e^{-i(\omega t-qx)}$. The velocity 
has only $x$-component (${\bf v}=\hat{{\bf x}}v$) and only 
diagonal component $\pi_{xx}$ gives a 
contribution into Eq. (\ref{24}). After differentiation 
with respect to $x$, Eq. (\ref{24}) takes the form:
\begin{equation}
-\partial_{t}^{2}\delta n+\frac{1}{m}\partial_{x}^{2}\delta P+
\frac{1}{m}\partial_{x}^{2}\pi_{xx}-\frac{en_{0}}{m}\partial_{x}^{2}\varphi=0
\label{27}
\end{equation}
Combining Eq. (\ref{23}) and Eqs. (\ref{25},\ref{26}), we express $\delta P$ 
and $\pi_{xx}$ in terms of the density fluctuations $\delta n$: 
\begin{equation}
\delta P=\frac{5P_{0}}{3n_{0}}\delta n,\quad 
\delta \pi_{xx}=\frac{4P_{0}}{3n_{0}}\delta n.  
\label{28}
\end{equation}
Substitution of (\ref{28}) into (\ref{27}) together with the Poisson equation
(\ref{9}) gives the dispersion of the plasma waves:
$$
\omega^{2}(q)=\omega_{p}^{2}+v_{0}^{2}q^{2}
$$
where
\begin{equation}
v_{0}^{2}=\frac{5P_{0}}{3mn_{0}}+\frac{4P_{0}}{3mn_{0}}=
\frac{3P_{0}}{mn_{0}}\equiv <v_{p}^{2}>.  
\label{29}
\end{equation}
The first contribution in $v_{0}^{2}$ equals to the square of the sound
velocity $v_{s}^{2}$ (the result of the ordinary hydrodynamics) 
and comes from the fluctuations of pressure $\delta P$.
The second one arises from the traceless part of the stress tensor $\pi _{ij}
$. The sum gives a correct coefficient which in case of a degenerate Fermi
gas reduces to the well known result $\frac{3}{5}v_{F}^{2}$. It is
straightforward to show that taking into account the third and fourth
moments leads to the correct plasmon dispersion up to $q^{4}.$
 
In the low frequency (or static) limit an expression for collision integral
is needed. We take $I_{ij}$ in a Krook-Bhatnager-Gross (KBG) approximation
(see, i.e. \cite{Rukh}): 
\begin{equation}
I_{{\bf p}}[f_{{\bf p}}]=
-\nu \left( f_{{\bf p}}({\bf r},t)-f_{{\bf p}}^{F}({\bf r},t)\right) ,  
\label{30}
\end{equation}
where $f_{{\bf p}}^{F}$ is a local equilibrium Fermi function with
position-dependent velocity, chemical potential 
$\mu ({\bf r},t)$ and temperature $T({\bf r},t)$. To satisfy the
general properties (\ref{8}) the function $f_{{\bf p}}^{F}$ should
give the same values of velocity ${\bf v}({\bf r},t)$, density
$n({\bf r},t)$ and pressure $P({\bf r},t)$, as the exact
distribution function $f_{{\bf p}}({\bf r},t)$.
 
The second moment of (\ref{30}) is equal to
\begin{equation}
I_{ij}=-\nu \left( P_{ij}-\delta_{ij}P_{F}(n,T)\right) ,  
\label{31}
\end{equation}
where $P_{F}(n,T)$ is the pressure of the Fermi gas with the distribution
function $f_{{\bf p}}^{F}$. By definition, this pressure equals to the exact
pressure, $P_{F}(n,T)=P\equiv TrP_{ij}$ hence the expression in the
brackets in (\ref{31}) equals to $\pi_{ij}$. Consequently $I_{ij}$ can be
written as
\begin{eqnarray}
I_{ij} &=&-\nu \pi_{ij},  
\label{32} \\
P &=&P_{F}(n,T).  
\label{33}
\end{eqnarray}
Eqs. (\ref{4}), (\ref{20}-\ref{22}) with $I_{ij}$ from (\ref{32}) and $P$
from (\ref{33}) constitute a closed set of equations of the generalised
hydrodynamics in the second order of $\gamma $. Here Eq. (\ref{33}) does
not mean the assumption of the local equilibrium, but simply introduces the 
new variable $T({\bf r},t)$ instead of $P({\bf r},t)$. In the low frequency
limit the system is in a local equilibrium and
$T({\bf r},t)$ coincides with the real local temperature.
 
It is easy to see that Eqs.(\ref{20}-\ref{22},\ref{32},\ref{33}) 
transform into the common 
hydrodynamics in the limit $\omega/\nu \ll 1$. Indeed, in a zero order of 
$\omega /\nu $ Eq. (\ref{22}) leads to $\pi _{ij}=0$, and Eqs. (\ref{20}) and
(\ref{21}) become identical respectively to Euler equation and to the equation 
for the conservation of energy in an ideal liquid. In the first order of 
$\omega/\nu \ll 1$ Eq. (\ref{22}) has a solution 
\begin{equation}
\pi _{ij}=-\frac{P}{\nu }\left( \nabla _{i}v_{j}+
\nabla _{j}v_{i}-\frac{2}{3}\delta _{ij}\nabla _{k}v_{k}\right) ,  
\label{34}
\end{equation}
which is the viscosity tensor with a viscosity coefficient $\eta =P/\nu $. In
this case Eqs.(\ref{20}, \ref{21}) are equivalent to Navies-Stocks equation
and the equation for the energy conservation in a viscous liquid.
 
Though the formulas (\ref{32}, \ref{33}) were obtained using the KBG 
collision integral (\ref{30}), they are more general than the KBG
approximation itself. The reason is that in the low frequency range
where collisions are important, the second moment
of $I$ is always a linear function of $\pi_{ij}$ \cite{Klim}. The
coefficient $\nu $ can thus be considered as a phenomenological parameter,
which is connected to viscosity as $\eta =P/\nu $.
 
Yet the system (\ref{4}), (\ref{20}-\ref{22}) in a low frequency limit is
not fully equivalent to classical hydrodynamics since the thermal conductivity
contribution $\nabla^{2}T$ \cite{Landau} is missing in the
energy conservation equation. Without this contribution the correct static
limit $\nabla T({\bf r})=0$ cannot be recovered. The term $\nabla^{2}T$
corresponds to the correction of the fourth order of $\gamma$ in the
continuity equation. Thus to include the thermal conductivity one has to
consider the third and the fourth moments and neglect the fifth moment. This
calculation will be presented in a more detailed publication.
 
In the case of a dense degenerate Fermi gas $T/E_{F}\ll 1$ ($E_{F}$ being the
local Fermi energy) it is unnecessary to consider the third and the fourth
moments. At $T=0$ Eq. (\ref{33}) reduces to 
$P=\frac{1}{5m}(3\pi^{2})^{2/3}n^{5/3}$ and (\ref{21}) transforms 
into condition $\pi_{ij}\nabla _{j}v_{i}=0$. These two equations 
together with  the system of differential equations  
\begin{eqnarray}
&D_{t}&n +n\nabla _{k}v_{k}=0  
\label{35} \\
mn&D_{t}&v_{i} +\nabla_{i}P+\nabla_{j}\pi_{ij}-en\nabla_{i}\varphi =0,
\label{36} \\
&D_{t}&\pi_{ij} +\pi_{ij}\nabla_{k}v_{k}+\pi_{ik}\nabla_{k}v_{j}+\pi
_{kj}\nabla_{k}v_{i} \nonumber \\
&+&P\left( \nabla_{i}v_{j}+\nabla_{j}v_{i}-\frac{2}{3}\delta_{ij}\nabla
_{k}v_{k}\right) =-\nu \pi_{ij},  
\label{37}
\end{eqnarray}
provide a complete system of equations of the generalised hydrodynamics
for a degenerate Fermi gas.

It is interesting to note a relation of the linearised version of the
generalised hydrodynamics in the high frequency (collisionless) limit
(\ref{23}-\ref{26}) to the theory of elasticity. Let us introduce the
displacement vector ${\bf u}({\bf r},t)$ as $\delta n =-n_0\nabla{\bf u}$.
The continuity equation (\ref{23}) gives the usual relation between
velocity and displacement $\partial_t{\bf u}={\bf v}$. Introducing
the stress tensor $\sigma_{ij}=-\delta P \delta_{ij}-\pi_{ij}$ one
can rewrite Eq. (\ref{24}) as
\begin{equation}
mn_{0}\partial^2_{t}u_{i}-\nabla_{j}\sigma_{ij}-en_{0}\nabla_{i}\varphi=0.
\label{38}
\end{equation}
According to the Eqs.(\ref{25},\ref{26}), the relationship of the tensor 
$\sigma_{ij}$ to $u_i$ is the same as in the elasticity theory
\cite{Landau1}:
\begin{equation}
\sigma_{ij}= K\nabla_k u_k\delta_{ij}+\mu\left( \nabla_{i}u_{j}+
\nabla_{j}u_{i}-\frac{2}{3}\delta_{ij}\nabla_{k}u_{k}\right),
\label{39}
\end{equation} 
where the bulk modulus $K$ and the shear modulus $\mu$ of an electron gas are
$K=\frac{5}{3}P_0$ and $\mu = P_0$.

The inclusion of collisions violates this one-to-one 
correspondence to the elasticity theory. In this case the shear stress
tensor $\sigma_{ij}-\frac{1}{3}\delta_{ij}\sigma_{kk}=-\pi_{ij}$ should be 
determined from the equation:
\begin{equation} 
\partial_{t}\pi_{ij} +\mu\partial_t\left( \nabla_{i}u_{j}+
\nabla_{j}u_{i}-\frac{2}{3}\delta_{ij}\nabla_{k}u_{k}\right) =-\nu\pi_{ij},  
\label{40}
\end{equation}
which follows from (\ref{26}) and (\ref{32}).
As suggested in Ref.\cite{Conti}, one may interpret this equation as if
it describes the medium with the complex frequency-dependent elastic constants.
However, a correspondence to so called {\it highly viscous fluids} is 
more natural. These fluids behave as solids at short intervals of time, 
but as viscous liquids on the large time scale \cite{Landau1}.
The equation for a shear stress tensor of such fluids has been 
phenomenologically introduced by Maxwell (see \cite{Landau1}) and 
exactly coincides with Eq.(\ref{40}). Hence, in a linear approximation,
an electron gas behaves as a kind of highly viscous fluid.

The work of one of the authors (I.T.) was supported by the Alexander von
Humboldt Foundation.

\end{document}